\documentclass[10pt,twocolumn,letterpaper]{article}

\usepackage[pagenumbers]{iccv} 

\usepackage{amssymb}
\usepackage[accsupp]{axessibility}  
\usepackage{algorithmic}
\usepackage{graphicx}
\usepackage{xcolor}
\usepackage{blindtext}
\usepackage{colortbl}  
\usepackage{newfloat} 
\usepackage{float} 
\usepackage{listings}  
\usepackage{multirow}  
\usepackage{booktabs}  
\usepackage{amsmath}  
\usepackage{bbding}  
\usepackage{mathrsfs}  
\usepackage{tabularx} 
\usepackage{pythonhighlight}  
\usepackage{flushend} 
\lstnewenvironment{PythonB}[1][]{\lstset{style=mypython, frame=none, breaklines=true, #1}}{}  
\usepackage[ruled,boxed]{algorithm2e}  
\usepackage{xcolor, soul}  
\sethlcolor{pink}  
\usepackage{pifont}  

\definecolor{iccvblue}{rgb}{0.21,0.49,0.74}
\definecolor{linkred}{rgb}{1.0,0,0}
\definecolor{refergreen}{rgb}{0,1.0,0}
\usepackage[pagebackref,breaklinks,colorlinks,
    linkcolor=linkred, 
    citecolor=iccvblue 
]{hyperref}
\usepackage{orcidlink}

\newcommand{\oldtask}{V2MR\xspace}
\newcommand{\task}{MGSV\xspace}
\newcommand{\framework}{MaDe\xspace}
\newcommand{\dataset}{\task-EC\xspace}

\newcommand{\specialcell}[2][c]{%
  \begin{tabular}[#1]{@{}c@{}}#2\end{tabular}}
\newcommand{\specialcellleft}[2][c]{%
  \begin{tabular}[#1]{@{}l@{}}#2\end{tabular}}

\definecolor{cGrey}{HTML}{F3F7F2} 

\newcommand{\delete}[1]{}

\newcommand{\envelope}{\ding{41}}
\newcommand{\pen}{\ding{45}}

\def\paperID{933} 
\def\confName{ICCV}
\def\confYear{2025}

\title{Music Grounding by Short Video}

\author{
    Zijie Xin$^{1 \text{\pen}}$\orcidlink{0000-0002-9220-8735} \and 
    Minquan Wang$^{2}$ \and 
    Jingyu Liu$^{1}$ \and 
    Ye Ma$^{2}$ \and 
    Quan Chen$^{2}$ \and 
    Peng Jiang$^{2}$ \and 
    Xirong Li$^{1 \text{\envelope}}$\orcidlink{0000-0002-0220-8310} \\
    ~\\
    $^1$~Renmin University of China \quad $^2$~Kuaishou Technology \\
    {\small \tt \url{https://github.com/xxayt/MGSV}}
}

\begin{document}
\twocolumn[{
\renewcommand\twocolumn[1][]{#1}%
\maketitle
\begin{center}
    \centering
    \captionsetup{type=figure}
    \includegraphics[width=1\textwidth]{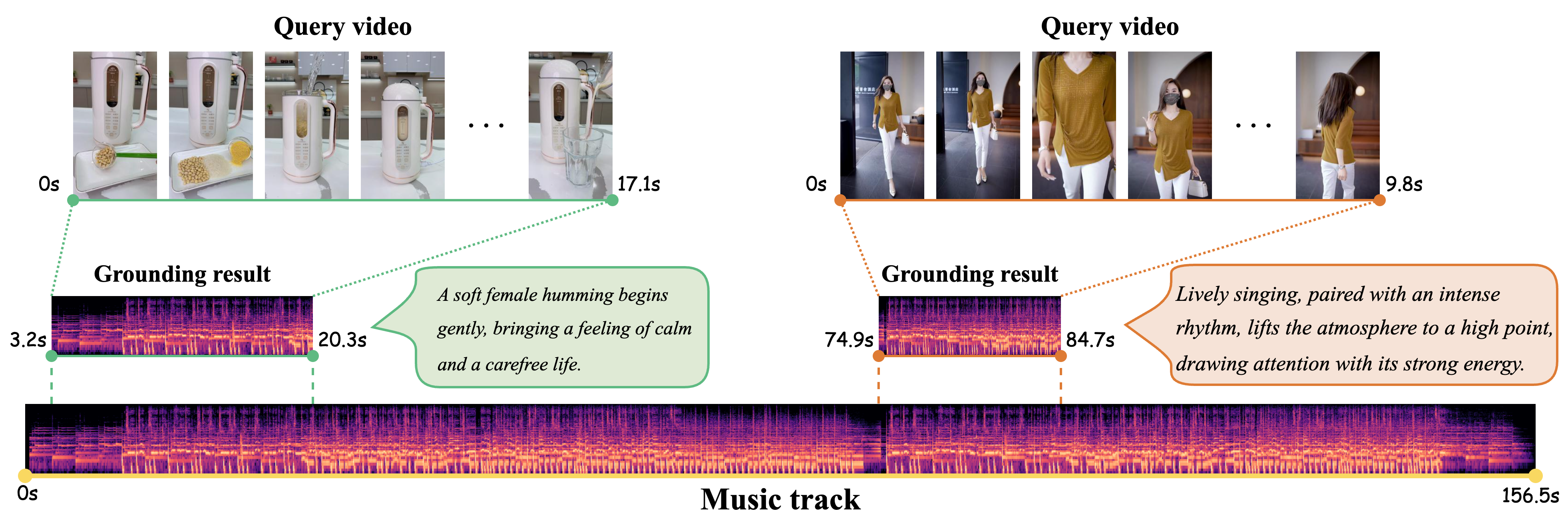}
    \captionof{figure}{\textbf{Music grounding by short video} (\task), aiming to localize within a music-track collection a \emph{music moment} that best serves as background music for the query video. Text in  callout boxes to the right of each music moment is for illustrative purposes only.}
    \label{fig:teaser}
\end{center}
}]

\renewcommand{\thefootnote}{\pen}
\footnotetext[1]{Work performed as an intern at Kuaishou. (xinzijie@ruc.edu.cn)}
\renewcommand{\thefootnote}{\envelope}
\footnotetext[2]{Corresponding author. (xirong@ruc.edu.cn)}

\renewcommand{\thefootnote}{\arabic{footnote}}
\setcounter{footnote}{0} 

\begin{abstract}
Adding proper background music helps complete a short video to be shared. Previous work tackles the task by video-to-music retrieval (\oldtask), aiming to find the most suitable music track from a collection to match the content of a given query video. In practice, however, music tracks are typically much longer than the query video, necessitating (\emph{manual}) trimming of the retrieved music to a shorter segment that matches the video duration. In order to bridge the gap between the practical need for \emph{music moment localization} and \oldtask, we propose a new task termed \underline{M}usic \underline{G}rounding by \underline{S}hort \underline{V}ideo (\task). To tackle the new task, we introduce a new benchmark, \dataset, which comprises a diverse set of 53k short videos associated with 35k different music moments from 4k unique music tracks. Furthermore, we develop a new baseline method, \framework, which performs both video-to-music \textbf{ma}tching and music moment \textbf{de}tection within a unified end-to-end deep network.  Extensive experiments on \dataset not only highlight the challenging nature of \task but also set \framework as a strong baseline.
\end{abstract} \label{sec:abs}
\section{Introduction} \label{sec:intro}

\emph{Music can name the unnameable and communicate the unknowable\footnote{Leonard Bernstein}}.
Adding appropriate background music (BGM) helps complete short videos, a dominant form of information dissemination online. To that end, 
current research focuses on video-to-music retrieval (V2MR), finding amidst a collection of music tracks the one best matching the content of a given query video \cite{li2019query,suris2022mvpt,mckee2023viml, Dong2024MuseChat}. In practice, music tracks (songs or instrumental music) are typically much longer than query videos (which usually last around 30 seconds). As a result, the retrieved music needs to be \textit{manually} cut to a shorter segment to match the video duration. In order to bridge the gap between the practical need and V2MR, we introduce in this paper a novel task called \textbf{M}usic \textbf{G}rounding by \textbf{S}hort \textbf{V}ideo (\task), aiming to localize within the music-track collection a \emph{music moment} that best serves as BGM for the query video, see \cref{fig:teaser}.

To address the new task, one might consider applying V2MR at a fine-grained segment level, with music tracks pre-trimmed into shorter segments. This approach is problematic due to the varying durations of videos. While \task is new, we observe in a broader context its conceptual resemblance to Video Grounding \cite{lei2021momentdetr,Jang2023EaTR,Xiao2024uvcom}, which aims to localize a moment in a specific video \wrt a given textual query. In particular, the query and target modalities in \task (Video Grounding) are video (text) and music (video), respectively. Thus, a Video Grounding method could, in principle, be repurposed for \task in a \emph{single-music} mode, assuming that the relevant music track for a query video is provided. When dealing with a collection of music tracks, V2MR is required to identify the relevant track. While combining existing Video Grounding and V2MR methods provides a good starting point, these methods were not originally designed for \task. By incorporating task-specific optimizations, we develop a stronger baseline that performs video-to-music \textbf{Ma}tching and music moment \textbf{De}tection (\textbf{MaDe}) in a single network.

\begin{figure*}[htbp!]
    \centering
    \begin{subfigure}{1.137\columnwidth}
      \includegraphics[width=1.0\columnwidth]{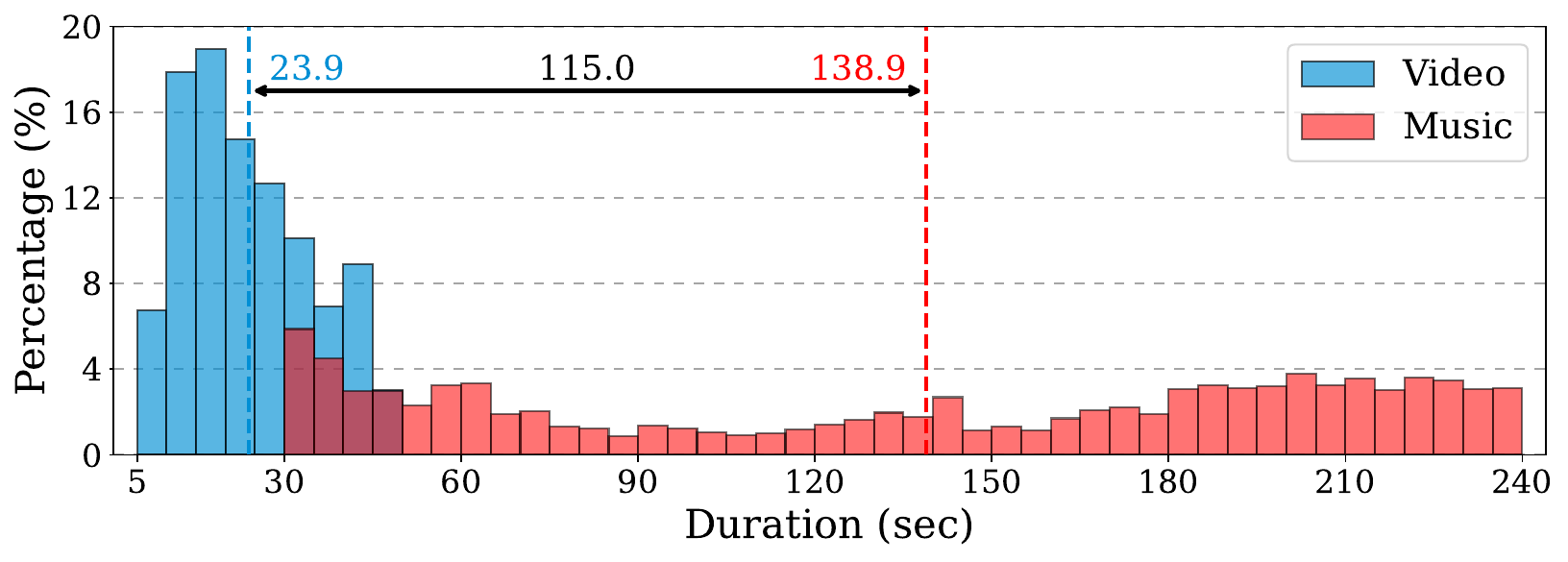}
      \caption{Distribution of video duration / music-track duration}
      \label{fig:duration}
    \end{subfigure}
    \hfill
    \begin{subfigure}{0.93\columnwidth}
      \includegraphics[width=1.0\columnwidth]{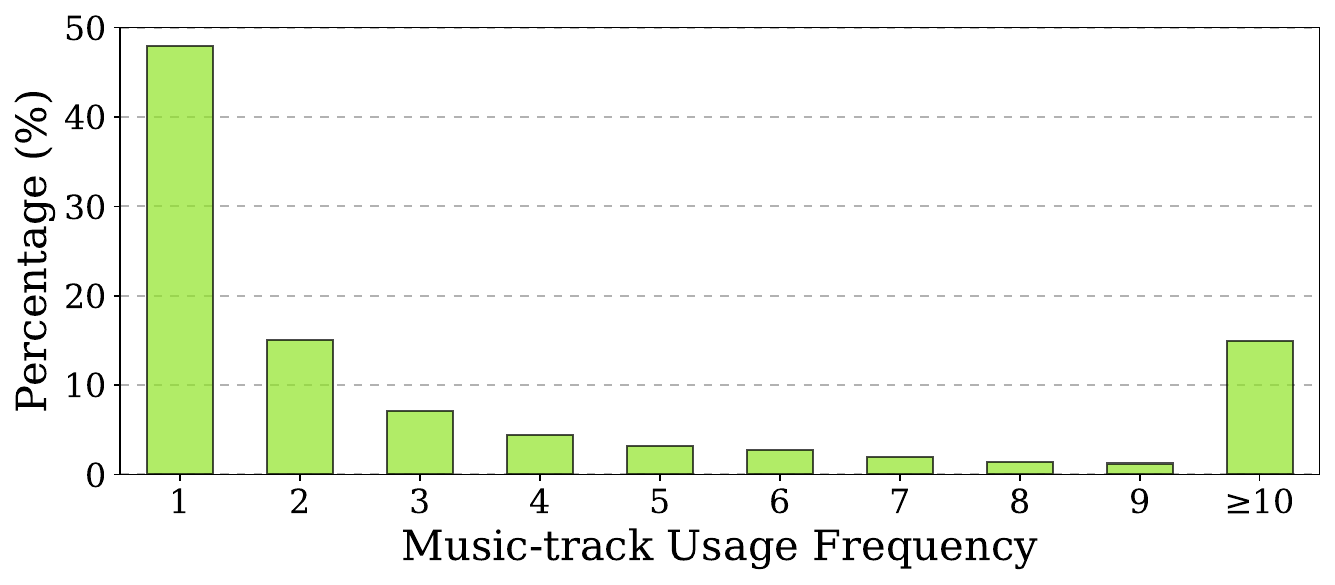}
      \caption{Usage frequency of music-track for video}
      \label{fig:usage_frequency}
    \end{subfigure}
    \\
    \colorbox{white!0}{\textcolor{white!0}{\rule{0.049\columnwidth}{2cm}}}  
    \begin{subfigure}{1.03\columnwidth}
      \includegraphics[width=1.0\columnwidth]{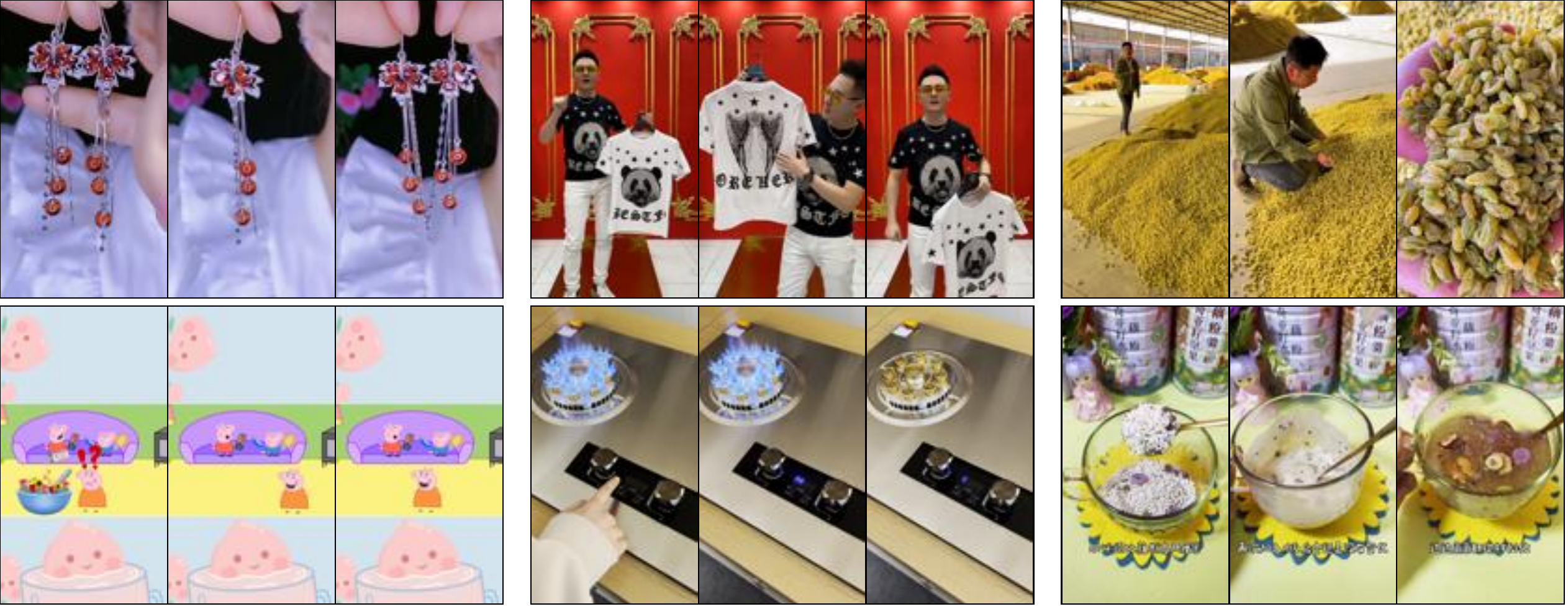}
      \caption{Snapshots of video samples}
      \label{fig:show_case}
    \end{subfigure}
    \hfill
    \begin{subfigure}{0.44\columnwidth}
      \includegraphics[width=1.0\columnwidth]{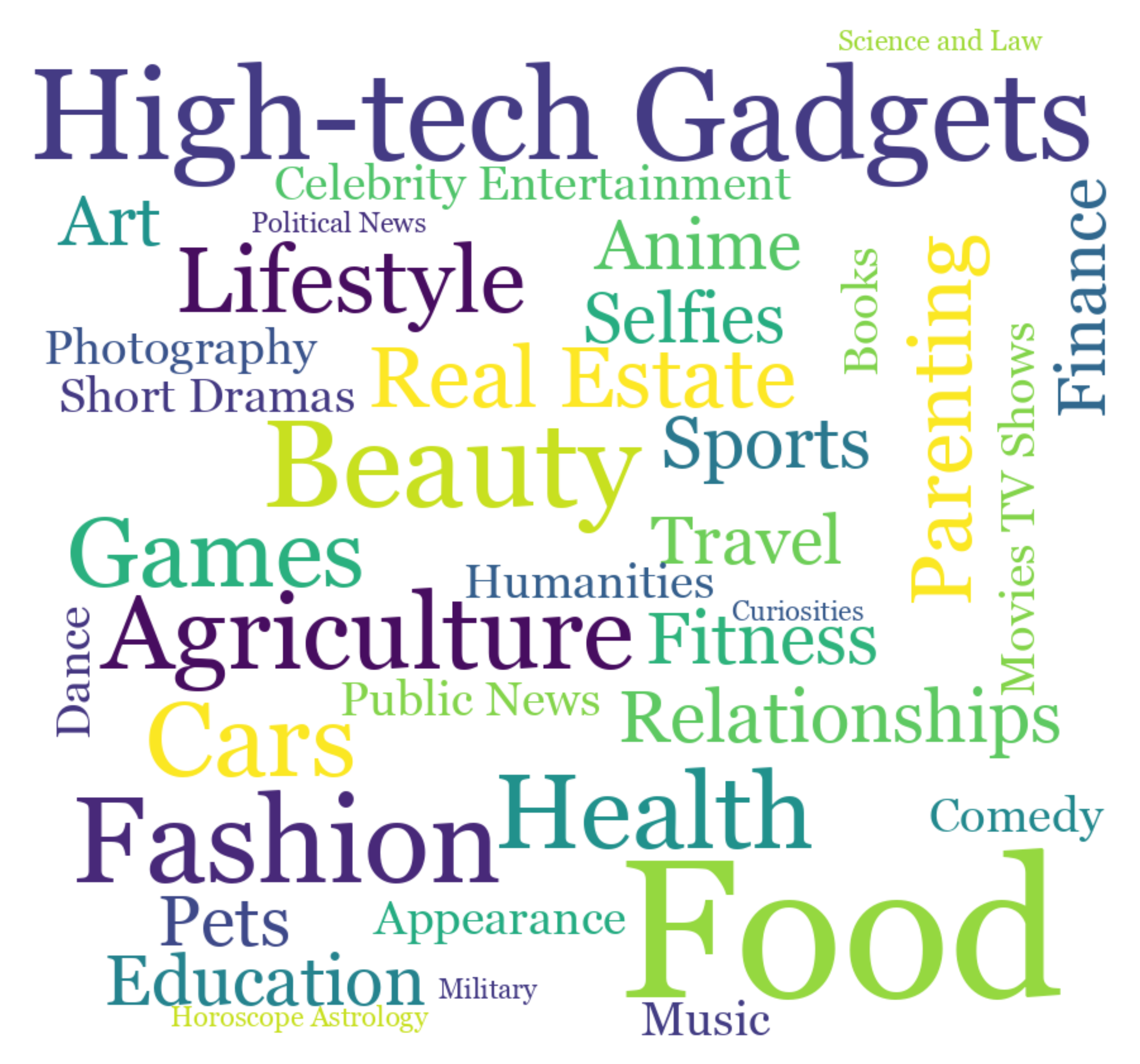}
      \caption{Video tag cloud}
      \label{fig:wordcloud}
    \end{subfigure}
    \hfill
    \begin{subfigure}{0.5\columnwidth}
      \includegraphics[width=1.0\columnwidth]{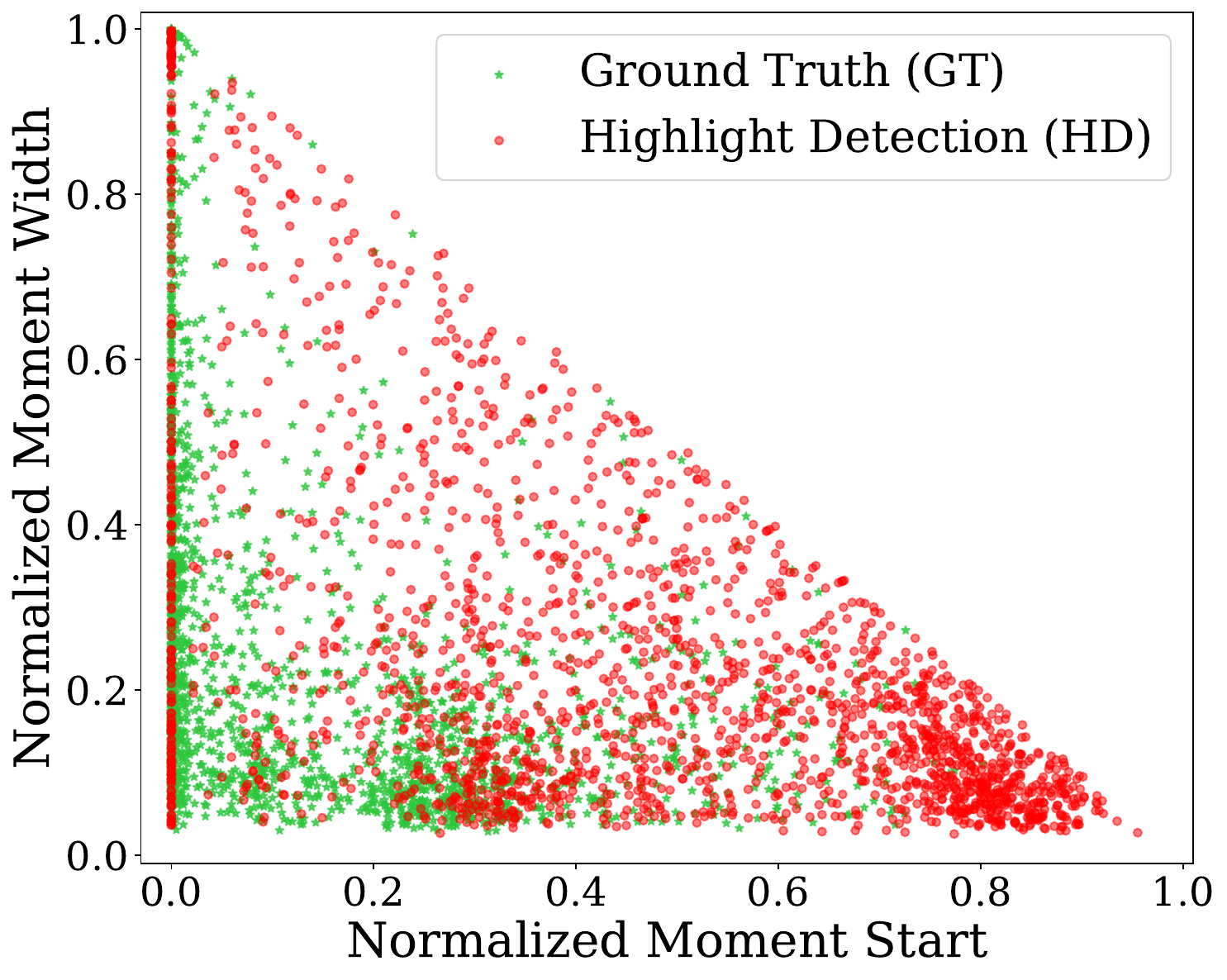}
      \caption{GT \textit{vs.} HD\cite{huang2018highlighter} moment position}
      \label{fig:moment_position}
    \end{subfigure}
    \caption{\textbf{Visualization of \dataset}. Note that the video tag cloud is merely to demonstrate the richness of the visual content. 
    }
    \label{fig:dataset-viz}
\end{figure*}

While few public datasets (mainly HIMV \cite{hong2018cbvmr}) exist for V2MR, each video in these datasets is paired with an equal-length music segment, yet the source tracks from which these segments were derived are absent. Therefore, they cannot be repurposed to \task. We construct a new dataset, \textbf{\dataset}, sourced from an \textbf{e}-\textbf{c}ommerce video creation platform. The availability of user-generated BGM editing logs on this platform allows us to reliably trace back to the source tracks. Noting that music selection for short videos involves inherent subjectivity, our work explores if an automated system can recommend music comparable to that made by skilled practitioners. As such,  this work is a starting point for answering the higher-level challenge of diversity. In sum, our main contributions are as follows:

\begin{itemize}
\item \textbf{Dataset}.  We introduce \dataset, a real-world dataset comprising 53k videos associated with 35k different music moments from 4k unique music tracks, see \cref{fig:dataset-viz}. This dataset supports the evaluation of methods for \task in two distinct modes: \emph{single-music} grounding (SmG) and \emph{music-set} grounding (MsG).

\item \textbf{Baseline}. We develop \framework, a unified end-to-end deep network that performs both video-to-music matching and music moment detection. \framework will be open-source.

\item \textbf{Evaluation}. We adapt and evaluate a diverse set of models, including six Video Grounding models (Moment-DETR \cite{lei2021momentdetr}, QD-DETR \cite{moon2023qddetr}, TR-DETR \cite{Sun2024TRDETR}, EaTR \cite{Jang2023EaTR}, and UVCOM \cite{Xiao2024uvcom}), one V2MR model (MVPt \cite{suris2022mvpt}), and one Video-corpus Moment Retrieval model (CONQUER \cite{hou2021conquer}). The evaluation highlights the challenging nature of \task and establishes \framework as a stronger baseline.
\end{itemize}

The rest of the paper is organized as follows. We discuss related work in \cref{sec:related}. The new dataset is introduced in \cref{sec:dataset}, followed by our solution in \cref{sec:method}. Experiments are presented in \cref{sec:eval}. We conclude the paper in \cref{sec:conc}.
\section{Related Work} \label{sec:related}

In developing our solution for \task, we draw inspiration from recent studies on Transformers-based Video Grounding and Video-to-Music Retrieval (V2MR).

\begin{table}[b!]
\centering
\renewcommand{\arraystretch}{1.2} 
\resizebox{1.0\linewidth}{!}{
\begin{tabular}{@{}lllcr@{}}
\toprule

\textbf{Model} & \specialcellleft{\textbf{Unimodal Feat.} \\ \textbf{Enhancement}} & \specialcellleft{\textbf{Multimodal} \\ \textbf{Feat. Fusion}} &  \specialcellleft{\textbf{Init. Content}\\ \textbf{{Token} $\phi_0$}} & \textbf{Decoder} \\
\midrule
Moment-DETR\cite{lei2021momentdetr} & FC$_{\times 2}$ & SA$_{\times 2}$ & 0 & SA-CA$_{\times 2}$ \\
\hline
QD-DETR\cite{moon2023qddetr} & FC$_{\times 2}$ & CA$_{\times 2}$+SA$_{\times 2}$ & 0 & SA-CA$_{\times 2}$ \\
QD-DETR+ & FC$_{\times 2}$ & SA$_{\times 2}$ & 0 & SA-CA$_{\times 2}$ \\
\hline
TR-DETR\cite{Sun2024TRDETR} & FC$_{\times 2}$ & \specialcellleft{VFR+CA$_{\times 2}$+SA$_{\times 2}$} & 0 & SA-CA$_{\times 2}$ \\ 
TR-DETR+ & FC$_{\times 2}$ & \specialcellleft{VFR+SA$_{\times 2}$} & 0 & SA-CA$_{\times 2}$ \\ 
\hline
EaTR\cite{Jang2023EaTR} & FC$_{\times 2}$ & SA$_{\times 3}$ & \specialcellleft{
Target-modality\\ features} & \specialcellleft{GF + \\SA-CA$_{\times 2}$} \\
\hline
UVCOM\cite{Xiao2024uvcom} & FC$_{\times 2}$ & \specialcellleft{Dual-CA$_{\times 3}$ + \\ DBIA+LRP+SA$_{\times 3}$} & \specialcellleft{
Target-modality\\ features} & SA-CA$_{\times 3}$ \\ 
UVCOM+ & FC$_{\times 2}$ & \specialcellleft{DBIA+LRP+SA$_{\times 3}$} & \specialcellleft{
Target-modality\\ features} & SA-CA$_{\times 3}$ \\ 
\hline
\textit{\framework} (this paper) & FC+SA & SA$_{\times 2}$ & \specialcellleft{
Query-modality\\ feature} & CA$_{\times 6}$ \\

\bottomrule
\end{tabular}
}
\caption{\textbf{Key elements in current DETR-based models for video grounding}. When adapting these models to the new \task task, we find that removing cross-attention (CA) blocks from their multimodal feature fusion module (where applicable) results in stronger baselines, which we denote with a ``+'' suffix.}
\label{table:vmr_summary}
\end{table}

\textbf{Video Grounding}, \emph{aka} video moment retrieval by natural language, involves localizing a moment within an untrimmed video that corresponds to a given natural language text.
While earlier methods for this task typically rely on pre-generated segments as candidate moments or moment proposals \cite{zhang2019man,Mithun_2019_CVPR,zhang2020learning}, more recent models, equipped with DETR-like decoding modules, directly regress temporal boundaries in a proposal-free manner \cite{lei2021momentdetr,moon2023qddetr,Sun2024TRDETR,Jang2023EaTR,Xiao2024uvcom}. Key elements of these models are summarized in \cref{table:vmr_summary}. For instance, consider Moment-DETR \cite{lei2021momentdetr}, the first work that adapted DETR \cite{carion2020detr} for the task. Given a sequence of visual tokens extracted from video frames and textual tokens from the input text, Moment-DETR first enhances the unimodal features and aligns their dimensions using fully connected (FC) layers. The enhanced tokens are then concatenated (\textcircled{c}) along their temporal axis and passed through two self-attention (SA) blocks for multimodal feature fusion. Subsequently, a cross-attention (CA) based decoder is employed, utilizing an array of 10 trainable query tokens as $Q$, with the previously fused tokens serving as $K$ and $V$ for moment location prediction. Note that each query token for a specific CA is obtained as the sum of a \emph{positional} token, which is shared across all CAs, and a \emph{content} token from the preceding CA. For its first CA, Moment-DETR initializes the content token with a ZERO vector. As shown in \cref{table:vmr_summary}, subsequent works such as QD-DETR \cite{moon2023qddetr}, TR-DETR \cite{Sun2024TRDETR}, and UVCOM \cite{Xiao2024uvcom} primarily focus on innovations in multimodal feature fusion to enhance the interaction between visual and textual tokens, largely through the use of cross-attention (CA) mechanisms. However, in our preliminary experiments, we find that models employing CA-based fusions are more prone to overfitting in the new task. Consequently, we remove CA from the fusion module.

There have been notable efforts to extend video grounding to a video-corpus scenario, known as Video Corpus Moment Retrieval (VCMR) \cite{hou2021conquer,zhang2021ReLoCLNet,Yoon2022SQuiDNet}. In particular, these methods first perform text-to-video retrieval to select the top-$k$ candidate videos from a given corpus. Subsequently, moment detection is conducted for each candidate video. It is worth noting that the above works implement moment detection in a proposal-based manner. For instance, SQuiDNet \cite{Yoon2022SQuiDNet} treats each sampled frame as a candidate point for the start or end timestamp of the moment to be detected, thereby formulating the task as a binary classification problem. Consequently, the temporal resolution of the detection result is limited by the frame sampling frequency.

\textbf{V2MR} aims to retrieve music tracks relevant to a given query video through video-to-music matching. Earlier approaches 
utilized two-branch MLPs for cross-modal feature alignment \cite{hong2018cbvmr,pretet2022segvmnet,chen2024VERIFIED}. In contrast, current methods are based on Transformers, employing self-attention (SA) mechanisms for long-range temporal modeling, see MVPt \cite{suris2022mvpt}. Building on MVPt as their backbone, ViML \cite{mckee2023viml} fuses text and video input representations to query music samples, while MuseChat \cite{Dong2024MuseChat} introduces an interactive V2MR system that enhances retrieval results through conversational interactions. Despite these advancements in video-music matching, music grounding is unexplored in these efforts.

\section{Dataset for \task: \dataset} \label{sec:dataset}
In order to develop and evaluate solutions for the new \task task, we build \dataset, a dataset consisting of 53k professionally made \textbf{E}-\textbf{c}ommerce videos, with their BGMs extracted from a set of 4k unique music tracks, see \cref{tab:new_datasets} and \cref{fig:dataset-viz}. We describe the data curation procedure in \cref{ssec:dataset_curation}, followed by the evaluation protocol in \cref{ssec:evaluation_criteria}.

\subsection{Dataset Curation} \label{ssec:dataset_curation}
\begin{table}[!b]
\centering
\vspace{-3pt}  
\renewcommand{\arraystretch}{1} 
\resizebox{1.0\linewidth}{!}{
\begin{tabular}{@{}lrrrrr@{}}
\toprule
\textbf{Split} & \textbf{Music tracks} & \textit{Duration}(s) & \textbf{Query videos} & \textit{Duration}(s) & \textbf{Moments} \\
\midrule
Total & 4,050 & 138.9$\pm$69.6 & 53,194 & 23.9$\pm$10.7 & 35,393 \\ [2pt]
\emph{Train} & 3,496 & 138.3$\pm$69.4 & 49,194 & 24.0$\pm$10.7 & 31,660 \\
\emph{Val.} & 2,000 & 139.6$\pm$70.0 & 2,000 & 22.8$\pm$10.8 & 2,000 \\
\emph{Test} & 2,000 & 139.9$\pm$70.1 & 2,000 & 22.6$\pm$10.7 & 2,000 \\
\bottomrule
\end{tabular}
}
\caption{\textbf{Overview of the \dataset dataset}. While the query videos have no overlap between train / val. / test, music tracks are partially shared across the data split. As the music tracks are meant for re-using, such a setup is practical. 
}
\label{tab:new_datasets}
\end{table}

\textbf{Raw Data Gathering}. 
With permission, we gathered videos posted on an E-commerce video creation platform in the period from Jan. to Sep. 2023, resulting in an initial set of 100k videos. The videos have a frame rate of 34 and a raw resolution of $1080\times 1920$. Each video is associated with a BGM editing log, indicating which music track was used and which part of the music was adopted as the BGM. A set of 9k music tracks was used in total. As the raw video set is quite diverse with varied data quality, automated data cleaning is performed as follows.

\textbf{Data Cleaning}.
Given the positive correlation between video content quality and view counts / follows, we empirically exclude videos that have no more than 200 view counts or 100 follows. Videos too short ($<$5 seconds) or too long ($>$50 seconds) are also removed. With the above cleaning procedure, we preserve 53,194 short videos accompanied with 35,393 different music moments from 4,050 unique music tracks. 
Among the tracks, songs and instrumental music are approximately in a 6:4 ratio.
As \cref{fig:duration} shows, the video duration is 23.9 seconds on average, whilst that of the music tracks is 138.9 seconds, indicating a noticeable duration gap between the query videos and the music tracks to be retrieved.
\cref{fig:moment_position} shows a large disparity between the ground-truth moment positions and those detected by a music highlight extractor\footnote{\href{https://github.com/remyhuang/pop-music-highlighter/}{https://github.com/remyhuang/pop-music-highlighter/}} \cite{huang2018highlighter}. These results clearly show the challenging nature of the new \task task.

\textbf{Data Split}. 
For reproducible research, we recommend a data split as follows. A set of 2k videos is randomly sampled to form a held-out test set,  while another random set of 2k videos is used as a validation set (for hyper-parameter tuning, model selection, \etc), with the remaining data used for training, see \cref{tab:new_datasets}.

\subsection{Evaluation Protocol} \label{ssec:evaluation_criteria}

Our benchmark supports two evaluation modes, \ie \emph{single-music}, wherein the music track relevant \wrt to a given query video is known in advance, and \emph{music-set}, wherein the relevant music track has to be retrieved from a given music track set. Mode-specific tasks and evaluation criteria are described as follows, see \cref{tab:mode}.

\begin{table}[htbp!]
\centering
\label{tab:mode}

\renewcommand{\arraystretch}{1} 
\resizebox{0.9\linewidth}{!}{
\begin{tabular}{@{}lll@{}}
\toprule

\textbf{Mode} & \textbf{(Sub-)Tasks} & \textbf{Metrics}  \\ 
\midrule
\textit{Single-music} & Grounding (SmG) & mIoU \\ [3pt]
\multirow{2}{*}{\textit{Music-set}} & Video-to-Music Retrieval (V2MR) & R$k$ \\
 & Grounding (MsG) & MoR$k$ \\
\bottomrule
\end{tabular}
}
\caption{\textbf{Evaluation modes, (sub-)tasks and metrics}.
}
\end{table}

\textbf{\textit{Single-music} Mode}.
In this mode, a user already knows in their mind which music track to use. A model for \task shall automatically detect and cut from the track an appropriate moment based on the query video. In order to evaluate the accuracy of single-music grounding (\textbf{SmG}), per query video we compute temporal Intersection over Union (\textbf{IoU}) between the predicted moment and the corresponding ground truth. Higher IoU is better. A mean IoU (\textbf{mIoU}) is obtained by averaging IoU scores over all query videos.

\textbf{\textit{Music-set} Mode}.
In contrast to the single-music mode, the music-set mode is much more challenging as which music track is relevant \wrt to the query video is unknown. Hence, the effectiveness of a model is jointly determined by its performance in two sub-tasks, \ie video-to-music retrieval (\textbf{V2MR}) for finding the relevant music track and music-set grounding (\textbf{MsG}) to localize the relevant moment. For V2MR, we adopt the popular Recall@$k$ (\textbf{R$k$}), \ie the percentage of query videos that have their corresponding music tracks ranked in the top-$k$ retrieval results ($k$=1, 5, 10). To evaluate MsG, we borrow IoU-conditioned R$k$ from \cite{hou2021conquer}, calculating Moment Recall (\textbf{MoR$k$}) as the percentage of query videos that have corresponding music moments (with IoU$>$0.7) recalled in the top-$k$ detected moments ($k$=1,10,100). For a fair comparison across models, we recommend generating the top-$k$ moments using a unified post-processing strategy: picking up one moment for each of the top-$k$ ranked music tracks.
\section{A Strong Baseline for \task: \framework}\label{sec:method}
\begin{figure*}[t!]
    \centering
    \includegraphics[width=\textwidth]{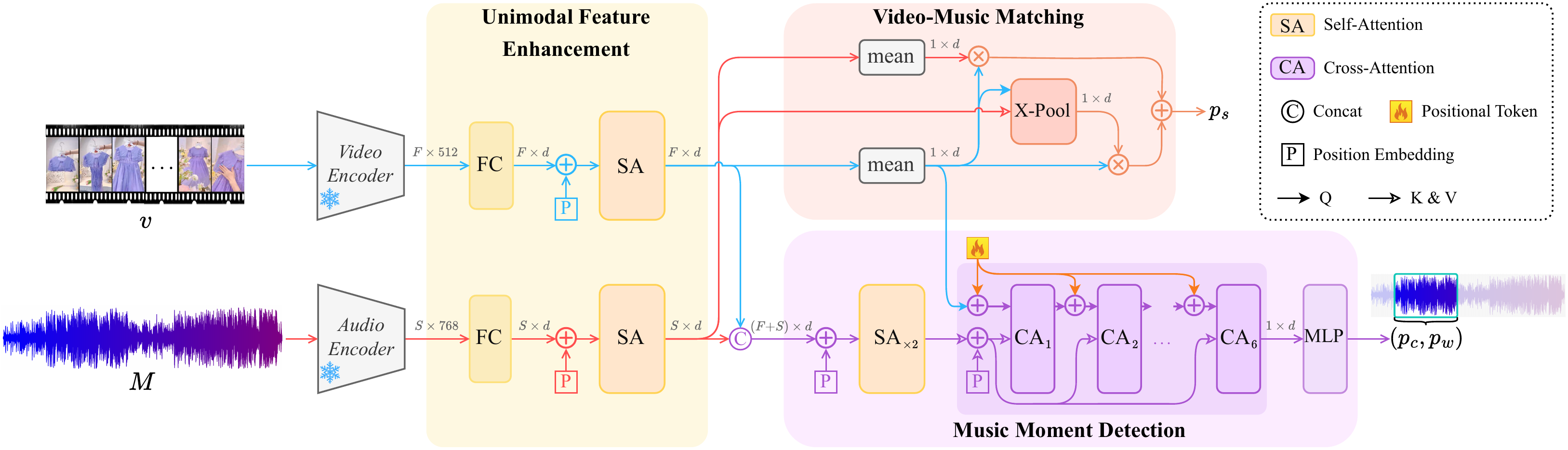}
    \caption{
    \textbf{Conceptual diagram of the proposed \framework network for \task}. At a high level, given a query video $v$ and a music track $M$ as multi-modal input, \framework yields a three-dimensional output ($p_s$, $p_c$, $p_w$), where $p_s$ measures the relevance of $M$ \wrt $v$, whilst $p_c$ / $p_w$ indicates the center / width of the detected music moment. The video is initially represented by a sequence of $F$ evenly sampled frames, and the music track as a sequence of $S$ partially overlapped Mel-spectrogram segments.  Pre-trained CLIP (ViT-B/32) and Audio Spectrogram Transformer (AST)  are used as our (weight-frozen) video and audio encoders, respectively. Each encoder is followed by an FC layer to compress frame / audio embeddings to a smaller size of $d$(=256) for cross-modal feature alignment and for parameter reduction. In the \emph{single-music} mode where $M$ is known to be relevant \wrt $v$, $p_s$ will be ignored. In the \emph{music-set} mode, we rank all candidate music tracks by their $p_s$ and accordingly select the moment detected from the top-ranked music track as the grounding result.
    }
    \label{fig:framework}
\end{figure*}

Given a query video $v$ and a specific music track $M$ as multi-modal input, we aim for a music grounding network $\mathcal{G}$ that can simultaneously estimate the relevance of $M$ \wrt $v$ and detect an appropriate music moment as the video's BGM. More formally, letting $p_s$ indicate the relevance score and $p_c$ / $p_w$ as the (normalized) temporal center / width of the moment, we express the network computation at a high level as:
\begin{equation} \label{eq:general}
(p_s, p_c, p_w) \leftarrow \mathcal{G}(v, M).
\end{equation}
One might question the necessity of predicting \( p_w \), as the moment length appears to be directly obtainable from the query video. We argue that enabling \(\mathcal{G}\) to predict \( p_w \) enhances its awareness of the moment boundaries, thereby improving its overall grounding capability. 
Note that in the \emph{single-music} mode where $M$ is known to be relevant \wrt $v$, $p_s$ will be ignored. As for the \emph{music-set} mode, all the candidate music tracks will be ranked by their $p_s$ and accordingly, the moment detected from the top-ranked music track will be chosen as the final prediction. Next, we elaborate the network architecture of $\mathcal{G}$ in \cref{ssec:network}, followed by the description of network training in \cref{ssec:training}.

\subsection{The \framework Network} \label{ssec:network}

The overall structure of our network is illustrated in \cref{fig:framework}. Conceptually, \framework consists of three modules, \ie 1) \textbf{multimodal embedding} (\cref{sssec:mm-embedding}) that converts the given query video and music track to a sequence of visual and audio tokens, respectively, 2) \textbf{video-music matching} (\cref{sssec:matching}) which estimates the relevance of the music track \wrt the video based on their holistic embeddings, and 3) \textbf{music moment detection} (\cref{sssec:detection}) that localizes within the music track the moment best fitting the video.

\subsubsection{Multimodal Embedding}\label{sssec:mm-embedding}

\textbf{Raw Feature Extraction}. 
Following the current practice of audiovisual analysis \cite{Girdhar2023imagebind,aaai25-VDSFX}, we adopt pre-trained CLIP (ViT-B/32) \cite{icml21-clip} and Audio Spectrogram Transformer (AST) \cite{gong2021ast} as our (weights-frozen) video and audio encoders, respectively. 
In particular, the query video is decoded at 1 FPS, yielding a sequence of $F$ frames resized to $256\times 256$. A 512-d CLS token is extracted per frame. 
As for the music track, we use AST's data preprocessing procedure to generate 
a sequence of $S$ partially overlapped 128-bin  Mel-spectrogram features. These Mel features are then fed into AST to obtain 768-d audio tokens $\{s_i\}_{i=1}^S$. 


\textbf{Unimodal Feature Enhancement}.  
For cross-modal embedding alignment and for parameter reduction, we place a fully connected (FC) layer after each encoder to compress the visual / audio tokens to a smaller size of $d$ (256). The compressed tokens are denoted as $\{\hat{f}_i\}_{i=1}^F$ and $\{\hat{s}_i\}_{i=1}^S$, respectively. Since the visual (audio) tokens are extracted independently per frame (segment), the inter-frame (inter-segment) connection along the temporal dimension is ignored. In order to enhance the tokens by temporal modeling, we feed them to a self-attention (SA) block \cite{vaswani2017attention}. 
Given $\{\tilde{f}_i\}$ and $\{\tilde{s}_i\}$ as temporally enhanced tokens, we express the  multi-modal embedding module more formally as:
\begin{equation} \label{eq:embedding}
\resizebox{.63\hsize}{!}{$
\left\{ \begin{array}{ll}
\{f_i\}_{i=1}^F & \leftarrow \mbox{ViT}(v, F),\\ [1pt]
\{\hat{f}_i\}_{i=1}^F & \leftarrow \mbox{FC}_{512\times d}(\{f_i\}_{i=1}^F), \\ [1pt]
\{\tilde{f}_i\}_{i=1}^F & \leftarrow \mbox{SA}(\{\hat{f}_i\}_{i=1}^F),\\ [3pt]
\{s_i\}_{i=1}^S & \leftarrow \mbox{AST}(M, S), \\ [1pt]
\{\hat{s}_i\}_{i=1}^S & \leftarrow \mbox{FC}_{768\times d}(\{s_i\}_{i=1}^S), \\ [1pt]
\{\tilde{s}_i\}_{i=1}^S & \leftarrow \mbox{SA}(\{\hat{s}_i\}_{i=1}^S).
\end{array} \right.
$}
\end{equation}



\subsubsection{Video-Music Matching} \label{sssec:matching}

For video-music matching, holistic (video-level / track-level) embeddings are required. To obtain such two embeddings, denoted as $h(v)$ and $h(M)$, a simple and commonly used operation is mean pooling over $\{\tilde{f}_i\}_{i=1}^F$ and $\{\tilde{s}_i\}_{i=1}^S$, respectively.
However, as shown in \cref{fig:duration}, the large disparity between short video duration and long music-track duration suggests that the music's relevance to the video is \emph{partial}. Hence, the mean-pooled embedding, denoted as $h_0(M)$, with  the $S$ segment embeddings treated equally, is suboptimal 
to reflect such partial relevance. 

We improve video-text matching by adopting X-Pool \cite{cvpr22-xpool} 
to obtain a \emph{video-conditioned} 
track embedding $h_1(M)$. 
Originally developed for video-text retrieval, X-Pool aggregates an array of local features with a cross-attention (CA) mechanism. In the current context, the video embedding $h(v)$, obtained by mean pooling, is used as $Q$, whilst $\{\hat{s}_i\}_{i=1}^S$ are used as $K$ and $V$. The attention weights generated by CA are used for weighted feature aggregation. The weights are computed based on the dot product between $Q$ and $K$ and thus reflect segment-wise relevance \wrt the video. Hence, music segments closer to the video will contribute more to $h_1(M)$.
The final video-music similarity $p_s$ is calculated as the sum of the cosine similarity ($cs$) between $h(v)$ and $h_0(M)$ and that of $h(v)$ and $h_1(M)$. In short, we have X-Pool enhanced video-music matching as:
\begin{equation} \label{eq:xpool}
\resizebox{.9\hsize}{!}{$
\left\{
\begin{array}{ll}
h(v) & \leftarrow \mbox{mean-pooling}(\{\tilde{f}_i\}_{i=1}^F),\\ [1pt]
h_0(M) & \leftarrow \mbox{mean-pooling}(\{\tilde{s}_i\}_{i=1}^S),\\ [1pt]
h_1(M) & \leftarrow \mbox{X-Pool}(h(v)~\mbox{as}~Q, \{\tilde{s}_i\}_{i=1}^S~\mbox{as}~K/V), \\ [1pt]
p_s & \leftarrow cs(h(v),h_0(M)) +cs(h(v),h_1(M)).
\end{array}
\right.
$}
\end{equation}

\subsubsection{Music Moment Detection} \label{sssec:detection}

In order to predict the moment localization ($p_c$ and $p_w$) based on the visual and audio tokens ($\{\tilde{f}_i\}_{i=1}^F$ and $\{\tilde{s}_i\}_{i=1}^S$), we adapt Moment-DETR \cite{lei2021momentdetr} as follows. Similar to Moment-DETR, we first fuse the tokens via concat (\textcircled{c}) and two SA blocks, producing 
$F+S$ modality-fused tokens $\{c_i\}_{i=1}^{F+S}$. In the subsequent decoding stage, $\{c_i\}$ are used as $K$ and $V$ input to each CA block, see \cref{fig:framework}. Note that in Moment-DETR, each CA processes an array of 10 composite query tokens, which are the sum of 10 learnable \emph{positional} tokens $P$ shared among the CAs and the same amount of \emph{content tokens} generated by the previous CA. In other words, given $\phi_k$ as the output of the $k$-th CA block, we have $P+\phi_{k-1}$ as $Q$. Concerning the choice of $\phi_0$, \ie the content token for the first CA, we propose to use the video embedding $h(v)$ instead of $\mathbf{0}$ in Moment-DETR. Such a simple tactic is important for music moment detection. Moreover, as we target at grounding the best music moment for a given video,  we reduce the number of query tokens from 10 to 1. As such, the SA preceding each CA in Moment-DETR and meant for processing the query token sequence is no longer needed. Hence, different from previous DETR-based models, our decoder has 6 CAs and 0 SA, see \cref{table:vmr_summary}.  Finally, we obtain $p_c$ and $p_w$ by feeding the output of the last CA $\phi_6$ to an MLP\footnote{$\mbox{FC}_{d\times d}\rightarrow \mbox{ReLU} \rightarrow \mbox{FC}_{d\times d} \rightarrow \mbox{ReLU} \rightarrow \mbox{FC}_{d\times 2} \rightarrow \mbox{sigmoid}$}. The detection module is expressed more formally as:
\begin{equation} \label{eq:mmd}
\resizebox{.9\hsize}{!}{$
\left
\{
\begin{array}{ll}
\{c_i\}_{i=1}^{F+S} & \leftarrow \mbox{SA}_{\times 2}(\{\tilde{f}_i\}_{i=1}^{F} \text{\textcircled{c}} \{\tilde{s}_j\}_{j=1}^S),\\ [1pt]
\phi_0 & \leftarrow h(v),  \\ [1pt]
\phi_k & \leftarrow \mbox{CA}_k(P+\phi_{k-1}~\mbox{as}~Q,  \{c_i\}_{i=1}^{F+S}~\mbox{as}~K/V), \\ [1pt]
(p_c,p_w) & \leftarrow \mbox{MLP}(\phi_6).
\end{array}
\right.
$}
\end{equation}

\subsection{Network Training} \label{ssec:training}

\textbf{Loss for Video-Music Matching}. We adopt a symmetric (video-to-music and music-to-video) InfoNCE loss, commonly used for training cross-modal matching networks \cite{icml21-clip,Chen2020simCLR,He2019moco,cvpr24-TeachCLIP}. Recall that our video-music similarity combines the mean-pooled embedding based similarity and the X-Pooled embedding based similarity, see \cref{eq:xpool}. To compute such a multi-similarity based loss, we see two options: a single loss calculated using the combined similarity or a joint loss with its components calculated separately using the individual similarities. Previous work on video-text retrieval \cite{tmm2021-sea} shows that the latter is better. We follow their recommendation, calculating the InfoNCE loss for each similarity and using their equal combination as the matching loss.
 
 

\textbf{Loss for Music Moment Detection}. 
In order to measure the temporal misalignment between the predicted moment and the ground truth, we adopt the loss from Moment-DETR \cite{lei2021momentdetr}, which comprises an L1 regression loss and a generalized Intersection over Union (IoU) loss. Note that for stable and fast training, the ground-truth moment center and width are normalized by the maximum music duration. 

The matching loss and the detection loss are equally combined. As shown in \cref{fig:framework}, except for the weights-frozen video and audio encoders, our network is end-to-end trained by minimizing the combined loss. 

\textbf{Implementation Details}. 
We use Transformers in their default setup \cite{vaswani2017attention}: sinusoidal positional encoding, 8 heads, and a dropout rate of 0.1. Trainable weights are initialized with Kaiming init \cite{He2015kaiminginit}, except for the detection module, which uses Xavier init \cite{Glorot2010xavier} as  Moment-DETR.
Our optimizer is Adam \cite{Kingma2014Adam} with an initial learning rate of 1e-4. Following CLIP, we employ a cosine schedule \cite{loshchilov2016sgdr} with a warm-up proportion of $0.02$.
The network is trained for 100 epochs with mini-batch size of 512. Our experimental environment is PyTorch 1.13.1 and NVIDIA 3090 GPUs. 

\section{Evaluation} \label{sec:eval}

\subsection{Single-music Grounding} \label{ssec:single_music_model}

\textbf{Baselines}. For a fair and reproducible comparison, we opt for SOTA video grounding models that are DETR-based and open-source. Accordingly, we collect the following five models and repurpose them for the single-music grounding (SmG) task: Moment-DETR\footnote{\href{https://github.com/jayleicn/moment_detr}{https://github.com/jayleicn/moment\_detr}} (NIPS21) \cite{lei2021momentdetr}, QD-DETR\footnote{\href{https://github.com/wjun0830/QD-DETR}{https://github.com/wjun0830/QD-DETR}} (CVPR23) \cite{moon2023qddetr}, EaTR\footnote{\href{https://github.com/jinhyunj/EaTR}{https://github.com/jinhyunj/EaTR}} (ICCV23) \cite{Jang2023EaTR}, TR-DETR\footnote{\href{https://github.com/mingyao1120/TR-DETR}{https://github.com/mingyao1120/TR-DETR}} (AAAI24) \cite{Sun2024TRDETR}, and UVCOM\footnote{\href{https://github.com/EasonXiao-888/UVCOM}{https://github.com/EasonXiao-888/UVCOM}} (CVPR24) \cite{Xiao2024uvcom}. These models are provided with the same raw features, the same training data and the same detection loss as \framework.

As shown in \cref{table:vmr_summary}, QD-DETR, TR-DETR and UVCOM employ CA blocks for multimodal feature fusion. We observe in our preliminary experiments that the CAs tend to cause these models to overfit to the training data. We term their CA-removed counterparts as QD-DETR+, TR-DETR+ and UVCOM+, respectively. In total, we have eight baselines for SmG. Note that they do not support video-music matching and are therefore inapplicable to MsG.

\begin{table}[t!]
\centering

\setlength{\tabcolsep}{1.5pt} 
\renewcommand{\arraystretch}{1.1} 
\resizebox{1\linewidth}{!}{
\begin{tabular}{@{}lcccrrrcrrr@{}}
\toprule

\multirow{2}{*}{\textbf{Model}} & \multirow{2}{*}{\specialcell{\textbf{\#Params} \\ (M)}} & \multicolumn{1}{c}{\specialcell{\textbf{SmG}}} & & \multicolumn{3}{c}{\specialcell{ \textbf{\oldtask}}} & & \multicolumn{3}{c}{\specialcell{\textbf{MsG}}}\\
\cmidrule{3-3} \cmidrule{5-7} \cmidrule{9-11}
& & \multicolumn{1}{c}{\textit{mIoU}} & & \textit{R1} & \textit{R5} & \textit{R10}  & & \textit{MoR1} & \textit{MoR10} & \textit{MoR100} \\
\midrule


\multicolumn{7}{@{}l}{\textit{Video Grounding re-purposed:}} \\

TR-DETR, AAAI'24 \cite{Sun2024TRDETR} & 7.8 & 0.393 & & -- & -- & --  & & -- & -- & -- \\
QD-DETR, CVPR'23 \cite{moon2023qddetr} & 6.9 & 0.423 & & -- & -- & --  & & -- & -- & -- \\

EaTR, ICCV'23 \cite{Jang2023EaTR} & 8.5 & 0.588 & & -- & -- & --  & & -- & -- & -- \\

Moment-DETR, NIPS'21 \cite{lei2021momentdetr} & 4.3 & 0.630 & & -- & -- & --  & & -- & -- & -- \\

TR-DETR+ & 6.2 & 0.630 & & -- & -- & --  & & -- & -- & -- \\

QD-DETR+ & 5.4 & 0.634 & & -- & -- & --  & & -- & -- & -- \\

UVCOM, CVPR'24 \cite{Xiao2024uvcom} & 14.5 & 0.652 & & -- & -- & --  & & -- & -- & -- \\
UVCOM+ & 12.9 & 0.661 & & -- & -- & --  & & -- & -- & -- \\
[3pt]

\multicolumn{2}{@{}l}{\textit{Video-to-Music Retrieval:}} \\
MVPt, CVPR'22 \cite{suris2022mvpt} & 3.6 & -- & & 2.4 & 6.8 & 9.4  & & -- & -- & -- \\
MVPt+ & 3.6 & -- & & 6.7 & 11.9 & 14.9 & & -- & -- & -- \\ 
[3pt]

\multicolumn{3}{@{}l}{\textit{Composite solution:}} \\
MVPt+ / UVCOM+   & 16.5 & 0.661 & & 6.7 & 11.9 & 14.9  & & 5.4 & 11.8 & 23.0 \\ [3pt]

\multicolumn{7}{@{}l}{\textit{Video Corpus Moment Retrieval re-purposed:}} \\
CONQUER, MM'21 \cite{hou2021conquer} & 39.4 & 0.572 & & 5.8 & 11.0 & 13.5  & & 4.4 & 9.6 & 18.4 \\  
[3pt]


\rowcolor{green!8}\framework (\emph{this paper}) & 10.5 & \textbf{0.722} &  & \textbf{8.8} & \textbf{16.3} & \textbf{19.8} &  & \textbf{8.3} & \textbf{17.6} & \textbf{30.7} \\
\bottomrule
\end{tabular}
}

\caption{\textbf{Overall results}. \#Params excludes the (weights-frozen) video / audio encoders.
}
\label{table:mmrv_all_baseline}
\end{table}

\textbf{Results}.
The performance of the baselines is shown in \cref{table:mmrv_all_baseline}. The superior performance of QD-DETR+ and TR-DETR+ compared to their original counterparts (mIoU 0.634 \emph{versus} 0.423 and 0.630 \emph{versus} 0.393) clearly reveals the negative impact of their CA-based fusion on the performance. By contrast, UVCOM (mIoU 0.652) is relatively stable and better. We attribute this result to UVCOM's dual CAs wherein the query (video) features and the target (music) features are each used as $Q$ respectively, making the query information better preserved during feature fusion. The higher mIoU of UVCOM+ (0.661) indicates that removing the dual CAs is also beneficial.
With a mIoU score of 0.722, our \framework clearly outperforms all the baselines.

\subsection{Music-set Grounding} \label{ssec:music_set_model}

\textbf{Baselines}. To tackle the MsG task, a straightforward solution is to combine UVCOM+ with an existing V2MR model. 
We implement MVPt (CVPR22) \cite{suris2022mvpt}, a popular V2MR baseline that has served as a backbone in subsequent works \cite{mckee2023viml, Dong2024MuseChat}. While the original MVPt adopts the CLS token from the last SA block as the music embedding, we find that mean pooling over the output of the last SA  yields better performance. We refer to this variant as MVPt+. By first using MVPt+ to find a set of candidate music tracks and then using UVCOM+ to detect moments within the candidate tracks, we obtain a composite solution ``MVPt+ / UVCOM+'' for MsG. In addition, we include CONQUER\footnote{\href{https://github.com/houzhijian/CONQUER}{https://github.com/houzhijian/CONQUER}} (MM21) \cite{hou2021conquer}, a moment-proposal based VCMR model, re-training it in the current setup.

\textbf{Results}. 
As \cref{table:mmrv_all_baseline} shows, the composition solution outperforms CONQUER in both SmG (mIoU 0.661 \emph{versus} 0.572) and MsG (MoR1 5.4 \emph{versus} 4.4) tasks, suggesting the advantage of the moment-proposal free solution in generating more accurate music grounding results. 
Compared to the above baselines, \framework achieves consistently superior results across all metrics (R1 8.8 and Mo-R1 8.3), thereby establishing itself as a strong baseline for \task. 

\begin{table}[t]
\centering
\setlength{\tabcolsep}{4pt} 
\renewcommand{\arraystretch}{1.1} 
\resizebox{1\linewidth}{!}{
\begin{tabular}{@{}lrccrrrcrrr@{}}
\toprule

\multirow{2}{*}{\textbf{Model}} & \multicolumn{1}{c}{\specialcell{\textbf{SmG}}} & & \multicolumn{3}{c}{\specialcell{ \textbf{\oldtask}}} & & \multicolumn{3}{c}{\specialcell{\textbf{MsG}}}\\
\cmidrule{2-2} \cmidrule{4-6} \cmidrule{8-10}
& \multicolumn{1}{c}{\textit{mIoU}} & & \textit{R1} & \textit{R5} & \textit{R10}  & & \textit{MoR1} & \textit{MoR10} & \textit{MoR100} \\

MVPt+ / UVCOM+ & $0.676_{\pm0.013}$ & & $7.0_{\pm0.4}$ & $12.5_{\pm0.7}$ & $15.5_{\pm0.5}$ & & $5.6_{\pm0.1}$ & $12.0_{\pm0.3}$ & $24.3_{\pm1.0}$ \\ [1pt]

\rowcolor{green!8}\framework & $0.743_{\pm0.017}$ &  & $9.0_{\pm0.5}$ & $17.3_{\pm0.7}$ & $21.1_{\pm1.0}$ &  & $8.4_{\pm0.3}$ & $18.0_{\pm0.5}$ & $32.7_{\pm1.4}$ \\
\bottomrule
\end{tabular}
}
\caption{\textbf{Average performance across three distinct data splits}.}
\label{table:more_data_split}
\end{table}

To verify the robustness of our findings across different data configurations, we conducted extra experiments using two newly generated random splits, following the same data partitioning protocol (\cref{ssec:dataset_curation}). The superior performance of MaDe compared to the best baseline remains, see \cref{table:more_data_split}.

\begin{figure}[htbp!]
    \centering
    \includegraphics[width=0.9\columnwidth]{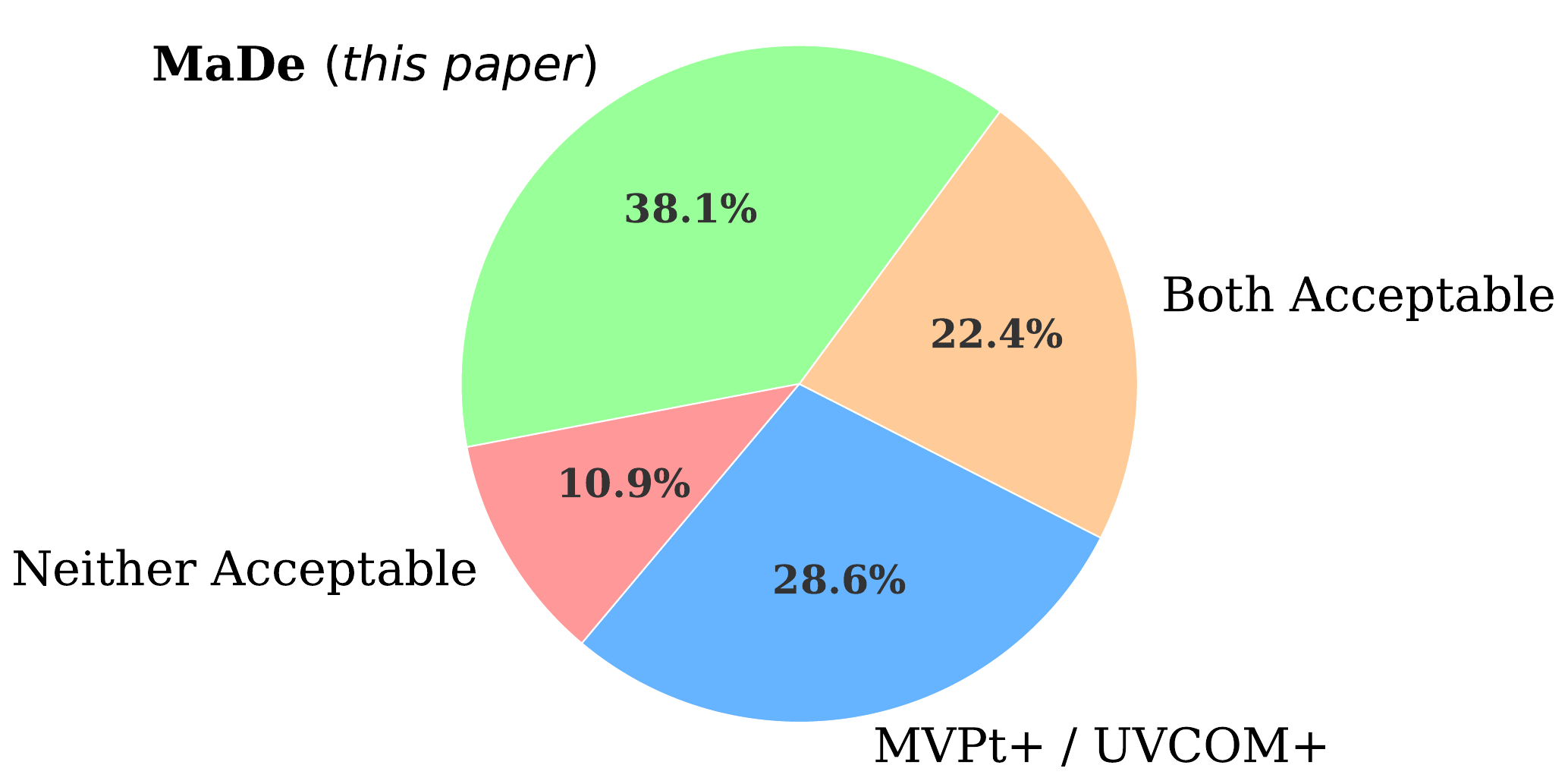}
    \caption{\textbf{Human evaluation results}.}
    \label{fig:user_study}
\end{figure}

\textbf{Human Evaluation}.
We further conduct a human evaluation on a random subset of 100 test videos to subjectively assess the quality of the MsG results. In particular, each video is individually paired with the music moment predicted by \framework and its counterpart by MVPt+ / UVCOM+. 
A subject is asked to watch the two resultant videos (labeled as Video A and Video B without revealing the method used) and, based on their overall visual and audio perception, select one of the following four options: \emph{A is better}, \emph{B is better}, \emph{neither acceptable}, or \emph{both acceptable}. Our evaluation team comprised 15 males and 5 females, all in their 20s and 30s, with frequent exposure to online short videos and music. \cref{fig:user_study} shows that \framework exceeds the composite solution by approximately 10\% (38.1\% \vs 28.6\%).

\begin{table}[tbp!]
\centering
\setlength{\tabcolsep}{3pt} 
\renewcommand{\arraystretch}{1.1} 
\resizebox{0.85\linewidth}{!}{
\begin{tabular}{@{}clrcrrr@{}}
\toprule

\multirow{2}{*}{\#} & \multirow{2}{*}{\textbf{Setup}} & \multicolumn{1}{c}{\textbf{SmG}} & & \multicolumn{3}{c}{\textbf{MsG}} \\
\cmidrule{3-3} \cmidrule{5-7}
 &  & \multicolumn{1}{c}{\textit{mIoU}} &  & \textit{MoR1} & \textit{MoR10} & \textit{MoR100} \\
\midrule

\rowcolor{green!8}0 & Full-setup & \textbf{0.722} &  & \textbf{8.3} & \textbf{17.6} & \textbf{30.7}  \\
[3pt]

\multicolumn{4}{@{}l}{\textit{Uni-modal Feature Enhancement:}} \\
1 & \textit{w/o} SA & 0.699 &  & 6.4 & 14.3 & 27.4 \\
2 & $\text{SA} \rightarrow \text{MLP}$ & 0.707 &  & 6.8 & 14.1 & 26.9 \\
[3pt]

\multicolumn{2}{@{}l}{\textit{Video-Music Matching:}} \\
3 & $cs(h(v),h_0(M))$ as $p_s$ & 0.708 &  & 7.1 & 15.8 & 29.1 \\
4 & $cs(h(v),h_1(M))$ as $p_s$ & 0.707 &  & 6.3 & 15.1 & 29.0 \\
5 & single loss & 0.715 &  & 7.5 & 16.8 & 29.2 \\
6 & $h_0(M)+h_1(M)$ & 0.716 &  & 6.9 & 16.0 & 28.3 \\
[3pt]

\multicolumn{2}{@{}l}{\textit{Music Moment Detection:}} \\
7 & \textit{w/o} $\text{SA}_{\times 2}$ & 0.705 & & 8.1 & 16.7 & 29.1 \\ 
8 & SA$_{\times 2}$ $\rightarrow$ CA & 0.697 &  & 7.1 & 16.4 & 29.1 \\
9 & $\mathbf{0}$ as $\phi_0$  & 0.709 &  & 7.4 & 16.4 & 29.0 \\ 
10 & $h_{0}(M)$ as $\phi_0$ & 0.719 &  & 8.0 & 17.4 & 30.4 \\
11 & $h_{1}(M)$ as $\phi_0$  & 0.718 &  & 7.5 & 16.9 & 29.5 \\
12 & \#Query-tokens: 1 $\rightarrow$ 10 & 0.716 & & 7.6 & 16.7 & 30.6 \\
13 & $(p_c, p_w) \rightarrow p_c$ & 0.706 &  & 7.3 & 16.5 & 29.0 \\

\bottomrule
\end{tabular}
}
\caption{\textbf{Ablation study of \framework}.
}
\label{table:mmrv_module_ablation}
\end{table}

\subsection{Ablation Study of \framework} \label{ssec:abl_exp}

For a better understanding of the superior performance of \framework, an ablation study is conducted as follows, with the results summarized in \cref{table:mmrv_module_ablation}.

\textbf{SA for Feature Enhancement}. Removing SA (Setup-1) or replacing it by an MLP (Setup-2) results in performance loss, with MoR1 decreased to 6.4 and 6.8, respectively.

\textbf{X-Pool for Video-Music Matching}. 
Using only the mean-pooled embeddings (Setup-3) results in a drop in Mo-R1 from 8.3 to 7.1. Using exclusively the X-Pooled embedding (Setup-4) leads to a more significant decline to 6.3. This result demonstrates that the two types of embeddings are complementary to each other. 
Regarding their joint exploitation, minimizing the joint loss outperforms using a single loss (Setup-5), and the combination based on similarity is superior to the feature-addition counterpart (Setup-6).

\textbf{SA for Multimodal Feature Fusion}. Removing SA$_{\times 2}$ (Setup-7) or replacing them with a CA (the music tokens as $Q$)(Setup-8) consistently results in performance loss.

\textbf{Choice of the Initial Content Token} $\phi_0$. We try three alternatives to $h(v)$, that is, $\mathbf{0}$ (Setup-9) and two target-modality features $h_0(M)$ (Setup-10) and $h_1(M)$ (Setup-11). The choice of using $h(v)$ as $\phi_0$ remains the best.

\textbf{Number of the Query Tokens}. We experimented with 10 query tokens, the default value in DETR and Moment-DETR, but observed lower performance (Setup-12).

\begin{table}[t!]
\centering
\renewcommand{\arraystretch}{1} 
\resizebox{0.6\linewidth}{!}{
\begin{tabular}{@{}lrrr@{}}
\toprule

\textbf{Num. CAs} & MoR$1$ & MoR$10$ & MoR$100$ \\
\midrule


1 & 7.6 & 16.7 & 28.3 \\
2 & 7.9 & 17.3 & 28.5 \\
3 & 7.4 & 16.0 & 27.2 \\
4 & 8.2 & 16.9 & 30.0 \\
5 & 7.3 & 16.2 & 29.2 \\
\rowcolor{green!8}6 & \textbf{8.3} & \textbf{17.6} & \textbf{30.7} \\


\bottomrule

\end{tabular}
}
\caption{\textbf{ 
Numbers of CAs for decoding}.}
\label{table:SA_CA_depth}
\end{table}

\textbf{Predicting $p_c$ only}. We modified \framework's detection head to predict only the moment center $p_c$, using the duration of the query video as the moment width (Setup-13). The resulting lower performance confirms the necessity of learning to predict temporal boundaries explicitly.

\textbf{Number of CAs for Decoding}. The influence of the number of CAs used for decoding is reported in \cref{table:SA_CA_depth}.

\begin{figure}[t]
    \centering
    \includegraphics[width=0.9\columnwidth]{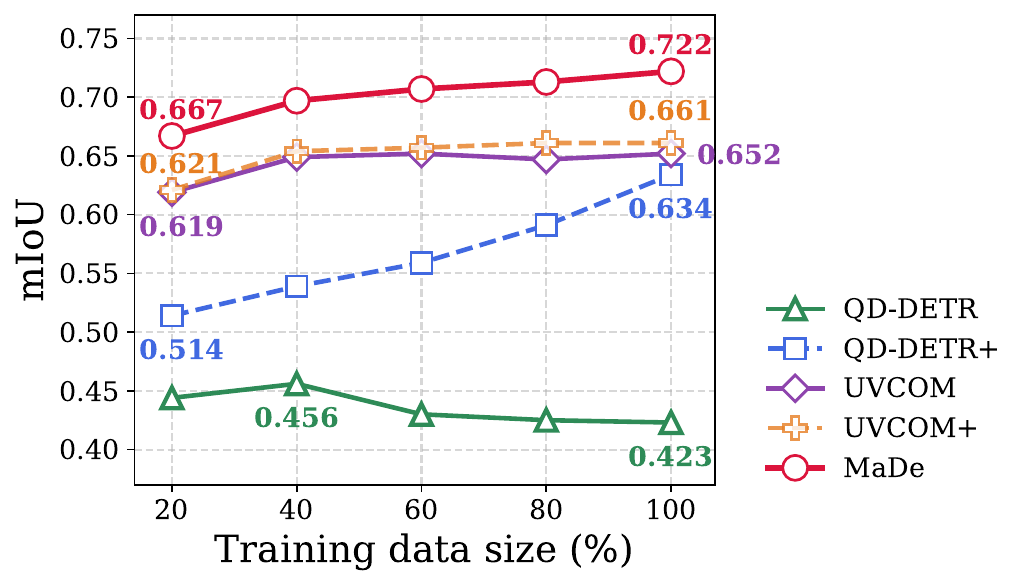}
    \caption{\textbf{SmG performance with varying training data sizes}.}
    \label{fig:SmG_train_data_usage}
\end{figure}

\textbf{CA for fusion}.
We conduct a controlled comparison between QD-DETR, UVCOM and their CA-free variants (QD-DETR+, UVCOM+) at varied training data scales. \cref{fig:SmG_train_data_usage} reveals an inverse relationship: QD-DETR declines with more data, while QD-DETR+ improves. This result suggests that CA-based multimodal feature fusion is the source of incompatibility with the MGSV task's demands.

\textbf{Efficiency Analysis}.
Given the raw visual / audio tokens precomputed, \framework performs music grounding for 512 video-music pairs in  0.09 seconds on average, see \cref{table:efficiency}.

\begin{table}[tbp!]
\centering
\renewcommand{\arraystretch}{1} 
\resizebox{0.85\linewidth}{!}{
\begin{tabular}{@{}lrr@{}}
\toprule

\textbf{Module} & \textbf{\#Params} (M) & \textbf{Time cost} (sec.) \\
\midrule
\emph{Feature enhancement} & 2.1 & 0.006 \\
\emph{Video-music matching} & 0.3 & 0.052 \\
\emph{Music moment detection} & 8.1 & 0.033 \\
Total & 10.5 & 0.091 \\
\bottomrule

\end{tabular}
}
\caption{\textbf{Inference efficiency per mini-batch}. Batch size: 512. Raw visual / audio tokens are 
precomputed and thus excluded.
}
\label{table:efficiency}
\end{table}



\section{Conclusions} \label{sec:conc}

In this paper, we have introduced Music Grounding by Short Video (\task) as a novel task and constructed MGSV-EC, a large-scale real-world dataset comprising 53k short videos associated with 35k different music moments from 4k unique tracks. Through extensive experiments on MGSV-EC, we draw the following conclusions. Among the adapted Video Grounding models for single-music grounding, UVCOM achieves the best performance. Removing cross-attention (CA) blocks from the multimodal feature fusion module consistently improves model performance. For the Video-to-Music Retrieval model MVPt, obtaining video and music embeddings by mean pooling outperforms its default option of using the CLS token. The superior performance of \framework on both single-music grounding and music-set grounding can be collectively attributed to the following optimizations: employing SA for unimodal feature enhancement, avoiding CA for multimodal feature fusion, jointly utilizing mean pooling and X-Pool for video-music matching, and using the video embedding as the initial content token for decoding. 
As such, \framework serves as a strong baseline for \task.

\textbf{Limitations}. Our study has clearly shown that CA-based multimodal feature fusion negatively impacts music grounding performance. The surface reason is that such a fusion mechanism tends to cause overfitting. However, the question of how video-music interaction needed for Music Grounding fundamentally differs from the text-video interaction required by Video Grounding remains open. How to customize music grounding with respect to the users' personal preferences is worthy of future research.

\paragraph{Acknowledgments} 
This research was supported by NSFC (No.62172420) and Kuaishou.
We thank Yingtong Liu, Yuchuan Deng, Yiyi Chen, and Bo Wang for valuable discussion and feedback on this research.

{
    \small
    \bibliographystyle{ieeenat_fullname}
    \bibliography{mgsv}
}

\end{document}